\documentclass[aps,pra,onecolumn]{revtex4-2}
\usepackage{graphicx}    
\usepackage{amsmath}

\begin{document}

\title{Collapse and revival of n-photon coherent states and n-photon squeezed coherent states}

\author{Moorad Alexanian}
\affiliation{Department of Physics and Physical Oceanography\\
University of North Carolina Wilmington\\ Wilmington, NC
28403-5606\\}

\date{\today}

\begin{abstract}
\noindent We introduce the set of $n$-photon coherent states and the set of $n$-photon squeezed coherent states and study their collapse and revival and compare their behavior with the collapse and revival of the corresponding  zero-photon coherent states  and zero-photon squeezed coherent states, respectively.  The set of $n$-photon squeezed coherent states ($n=0,1,2,\cdots$) forms a basis of the Hilbert state of photons and may be of some interest. We notice that the presence of a few extra photons in the $n$-photon states as compared to the corresponding zero-photon states makes a large difference in the behavior of their collapse and revival. This indicates the large effect that a few extra photons makes in the $n$-photon states as compared to the corresponding zero-photon states in their collapse and revival.
\end{abstract}

\maketitle {}


\section{Introduction}

 The Jaynes-Cummings model (JCM) \cite{JC63} of a two-level atomic system coupled to a single-mode radiation field is known to exhibit interesting optical phenomena, such as the collapse and revival of Rabi oscillations
 of the atomic coherence \cite{NSE81,B90,B91,SK93}. However, complex features may appear at the mesoscopic scale (few tens of photons) in the analytical analysis of the collapse and revival beyond the rotating wave approximation when the energy spectrum of the system is changed drastically \cite{ENS80,FS93,RBH01,FL09}. Recently, collapse and revival has been observed in a range of interaction times and photon numbers using slow circular Rydberg atoms interacting with a superconducting microwave cavity \cite{AGS19}. This experiment opens promising perspectives for the rapid generation and manipulation of non-classical states in cavity and circuit quantum electrodynamics.

 The present work is a sequel to two papers \cite{MA20,MA19} dealing with squeezed coherent states and quantum Rabi oscillations.  This paper is arranged as follows. In Sec. 2, we review the creation and annihilation operators that
adds or subtracts a photon from a coherent or a squeezed coherent state. In Sec. 3, we calculate the photon number variance in the $n$-photon squeezed coherent states, which is used in the text to minimize the quotient of the $n$-photon
number variance to the average number of photons. In  sec. 4, we review the dynamics governed by the Jaynes-Cummings model. In Sec. 5, we determine the collapse and revival of the $n$-photon coherent and the $n$-photon squeezed coherent states. Finally, Sec. 6 summarizes our results.

\section {Squeezed coherent photons}

Ia recent paper \cite{MA20}, the creation and annihilation operators $\hat{B}$ and $\hat{B^\dag}$, respectively, for the squeezed coherent photons are given by,  where $\hat{a}$ and $\hat{a}^\dag$ are the photon creation and annihilation operators,
\begin{equation}
\hat{B}= \hat{S}(\zeta)\hat{D}(\alpha)\hat{a}\hat{D}(-\alpha)\hat{S}(-\zeta)= \cosh(r)  \hat{a}+e^{i\varphi} \sinh(r) \hat{a}^\dag -\alpha
\end{equation}
and
\begin{equation}
\hat{B}^\dag=  \hat{S}(\zeta)\hat{D}(\alpha)\hat{a}^\dag \hat{D}(-\alpha)\hat{S}(-\zeta)=  e^{-i\varphi} \sinh(r)\hat{a} +\cosh(r)\hat{a}^\dag  -\alpha^*,
\end{equation}
with inverses
\begin{equation}
\hat{a}=\hat{D}(-\alpha)\hat{S}(-\zeta)\hat{B}\hat{S}(\zeta)\hat{D}(\alpha)=\cosh(r)\hat{B}-e^{i\varphi} \sinh(r) \hat{B}^\dag +\alpha\cosh(r)-\alpha^*e^{i\varphi} \sinh(r)
\end{equation}
and
\begin{equation}
\hat{a}^\dag =\hat{D}(-\alpha)\hat{S}(-\zeta)\hat{B}^\dag \hat{S}(\zeta)\hat{D}(\alpha)= -e^{-i\varphi} \sinh(r) \hat{B}+ \cosh(r) \hat{B}^\dag +\alpha^*\cosh(r)-\alpha e^{-i\varphi} \sinh(r),
\end{equation}
where
\begin{equation}
\hat{D}(\alpha)= \exp{(\alpha \hat{a}^\dag-\alpha^* \hat{a})}
\end{equation}
is the Glauber displacement operator with $\alpha=|\alpha|\exp{(i\theta)}$
and
\begin{equation}
\hat{S}(\zeta) = \exp{(-\frac{\zeta}{2}\hat{a}^{\dag 2}  +\frac{\zeta^*}{2}\hat{a}^{2})}
\end{equation}
is the squeezing operator with $\zeta =r\exp{(i\varphi)}$. Note that $[\hat{B},\hat{B}^\dag]=1$ follows from $[\hat{a},\hat{a}^\dag]=1$.

One can generate a squeezed coherent state from the vacuum as follows
\begin{equation}
|\zeta,\alpha,0\rangle \equiv \hat{S}(\zeta)) \hat{D}(\alpha)|0\rangle,
\end{equation}
where the state $|\zeta,\alpha,0\rangle$ is the vacuum state since $\hat{B}|\zeta,\alpha,0\rangle=0$.
Consider the state
\begin{equation}
\frac{(\hat{B}^\dag)^n}{\sqrt{n!}}|\zeta,\alpha,0\rangle=\hat{S}(\zeta)) \hat{D}(\alpha)\frac{(\hat{a}^\dag)^n}{\sqrt{n!}}|0\rangle \equiv|\zeta,\alpha,n\rangle,
\end{equation}
where the state $|\zeta,\alpha,n\rangle$ is called an $n$-photon squeezed coherent state with the ordinary squeezed coherent state $|\zeta,\alpha,0\rangle$ being the 0-photon coherent state. It follows from (1)-(4) and (8) that
\begin{align}
\begin{split}
  \hat{B}|\zeta,\alpha,n\rangle  & =\sqrt{n}|\zeta,\alpha,n-1\rangle\\
 \hat{B}^\dag|\zeta,\alpha,n\rangle & =\sqrt{n+1}|\zeta,\alpha,n+1\rangle\\
  \hat{B}^\dag \hat{B}|\zeta,\alpha,n\rangle &  =n|\zeta,\alpha,n\rangle.
\end{split}
\end{align}

The $n$-photon coherent state can be explicitly found with the aid of the binomial theorem and so
\begin{equation}
|0,\alpha,n\rangle=\frac{(\hat{a}^\dag -\alpha^*)^n}{\sqrt{n!}}|0,\alpha,0\rangle = \frac{(-\alpha^*)^n}{\sqrt{n!}} e^{-(|\alpha|^2/2)}  \sum_{k=0}^{\infty} \frac{\alpha^k}{\sqrt{k!}} \hspace{0.05in}{}_{0}F_{2}(0;-n,-k;-1/|\alpha|^2)|k\rangle, \hspace{0.2in} (n=1,2,\cdots)
\end{equation}
with the coherent state
\begin{equation}
|0,\alpha,0 \rangle =\exp(-|\alpha|^2/2)\sum_{n=0}^{\infty} \frac{\alpha^n}{\sqrt{n!}} |n\rangle
\end{equation}
and the generalized hypergeometric series \cite{GR80}
\begin{equation}
{}_{0}F_{2}(0;-n,-k;z)=\sum_{j=0}^{\infty}\frac{\Gamma(n+1)\Gamma(k+1)}{\Gamma(n+1-j)\Gamma(k+1-j)}\frac{z^j}{j!},
\end{equation}
which reduces to a polynomial since both $n$ and $k$ are positive integers.

Similarly, one can generate the more complex state $(n+1)$-photon squeezed coherent state from knowledge of the $n$-photon squeezed coherent state as follows
\begin{equation}
|\zeta,\alpha,n+1\rangle=\frac{\hat{B}^\dag}{\sqrt{n+1}}|\zeta,\alpha,n\rangle =\sum_{k=0}^{\infty}c_{k}^{(n+1)}|k\rangle
\end{equation}
and so
\begin{equation}
c_{k}^{(n+1)}=\frac{1}{\sqrt{n+1}}\big{(}\sqrt{k+1} \sinh(r)e^{-i\varphi} c_{k+1}^{(n)}+\sqrt{k} \cosh(r) c_{k-1}^{(n)} -\alpha^{*} c_{k}^{(n)}\Big{)}.
\end{equation}

For instance , one can generate the two-photon squeezed coherent state $|\zeta,\alpha,2\rangle$ from the one-photon squeezed coherent state  $ |\zeta,\alpha,1\rangle$ since
\begin{align}
\begin{split}
|\zeta,\alpha,1\rangle &=\sum_{k=0}^{\infty}\frac{e^{-(|\alpha|^2-e^{-i\varphi}\alpha^2\tanh{r})/2}}{\sqrt{2^k k! \cosh(r)}}\Big{(}-\frac{e^{-i\varphi/2}}{\sqrt{2}\cosh(r)\tanh^{1/2} (r)}  H_{k+1}(\frac{\alpha e^{-i\varphi/2}}{\sqrt{2\sinh(r)\cosh(r)}})\\  &+(\frac{\alpha e^{-i\varphi}}{\tanh(r)}   -\alpha^*)H_{k}(\frac{\alpha e^{-i\varphi/2} }{\sqrt{2\sinh(r)\cosh(r)}})\Big{)},
\end{split}
\end{align}
where $H_{k}(x)$ is the Hermite polynomial \cite{GR80} and so

\begin{align}
\begin{split}
|\zeta,\alpha,2\rangle &= \sum_{k=0}^{\infty}\frac{\tanh^{k/2}(r) e^{-a^2(1-\tanh(r))/2}}{\sqrt{2^{k+1}k!\cosh(r)}}\Big{(}-\frac{a\tanh(r)}{\sqrt{2\sinh(r)\cosh(r)}}H_{k+1}(\frac{a}{\sqrt{2\sinh(r)\cosh(r)}})\\
&+\big{(}-\frac{k}{\sinh(r)\cosh(r)}+\tanh(r) +a^2(1-\frac{1}{\tanh(r)})^2 \big{)} H_{k}(\frac{a}{\sqrt{2\sinh(r)\cosh(r)}}\Big{)},
\end{split}
\end{align}
where we have chosen $\varphi=2\theta$ and discarded an overall phase factor.

The set $\big{\{}|\zeta,\alpha,n\rangle\}$, $n=0,1,2, \cdots$, forms a basis since
\begin{equation}
\langle n',\alpha,\zeta  |\zeta,\alpha,n\rangle = \langle n'| \hat{D}(-\alpha)\hat{S}(-\zeta)  \hat{S}(\zeta)\hat{D}(\alpha) |n\rangle = \delta_{n'n}
\end{equation}
and
\begin{equation}
\sum_{n=0}^{\infty}|\zeta,\alpha,n\rangle \langle n,\alpha,\zeta| = \sum_{n=0}^{\infty}\hat{S}(\zeta))\hat{D}(\alpha) |n\rangle \langle n| \hat{D}(-\alpha)\hat{S}(-\zeta) =1.
\end{equation}
with the aid of $\langle n'|n\rangle= \delta_{n'n}$ and $\sum_{n=0}^{\infty} |n\rangle\langle n|=1$. Accordingly, the set is complete and orthonormal. Finally, one has for the mean number of photons in the $n$-photon squeezed
coherent state that
\begin{equation}
\langle n,\alpha,\zeta|\hat{a}^\dag \hat{a}|\zeta,\alpha,n\rangle =(n+1) \sinh^2(r)  +n\cosh^2(r) +|\alpha e^{-i\varphi/2}\cosh(r)-\alpha^* e^{i\varphi/2} \sinh(r) |^2.
\end{equation}

\section{photon variance}

The photon variance in the $n$-photon squeezed coherent state is obtained with the aid first of
\begin{equation}
\langle n,\alpha,\zeta|(\hat{a}^\dag \hat{a})^2|\zeta,\alpha,n\rangle=\sum_{n'=0}^{\infty}   \langle n,\alpha,\zeta|\hat{a}^\dag \hat{a}|\zeta,\alpha,n'\rangle \langle n',\alpha,\zeta|\hat{a}^\dag \hat{a}|\zeta,\alpha,n\rangle,
\end{equation}
where we have used the completeness relation (18). One obtains, with the aid of (3), (4), and (9), that
\begin{equation}
\langle n',\alpha,\zeta|\hat{a}^\dag \hat{a}|\zeta,\alpha,n\rangle =A\delta_{n'n} + B\delta_{n'n+1}+C\delta_{n'n-1}D\delta_{n'n+2} +E\delta_{n'n-2},
\end{equation}
where
\begin{align}
\begin{split}
A =& (n+1) \sinh^2(r)  +n\cosh^2(r) +|w|^2,\\
B=& \sqrt{n+1} \big{(} w^*\cosh(r)-we^{i\varphi}\sinh(r)\big{)},\\
C  =& \sqrt{n}\big{(}  w\cosh(r)-w^*e^{-i\varphi}\sinh(r)\big{)},\\
D =&- e^{-i\varphi}\sqrt{(n+1)(n+2)}\sinh(r) \cosh(r),\\
E =&-e^{-i\varphi}\sqrt{n(n-1)}\sinh(r) \cosh(r),
\end{split}
\end{align}
and $w=\alpha^*\cosh(r)-\alpha e^{-i\varphi}\sinh(r)$. Therefor, we have, for $n\geq 2$, that
\begin{equation}
\langle n,\alpha,\zeta|(\hat{a}^\dag \hat{a})^2|\zeta,\alpha,n\rangle = |A|^2 + |B|^2 +|C|^2 +|D|^2 +|E|^2
\end{equation}
with the variance
\begin{equation}
\Delta n^2= \langle \hat{n}^2\rangle -\langle\hat{n}\rangle^2 = |B|^2 +|C|^2 +|D|^2 +|E|^2.
\end{equation}
 where   $\hat{n} =\hat{a}^\dag \hat{a}$.

\section{Jaynes-Cummings model}

We suppose the dynamics is governed by the Jaynes-Cummings model with Hamiltonian \cite{JC63}
\begin{equation}
\hat{H}_{I}= \frac{\hbar\Delta}{2}\hat{\sigma}_{3}-i\hbar\lambda(\hat{\sigma}_{+}\hat{a}-\hat{a}^\dag \hat{\sigma}_{-}),
\end{equation}
where $\hat{N}=\hat{a}^\dag\hat{a}+\hat{\sigma}_{+}\hat{\sigma}_{-}$ is a constant of the motion and represents the number of quanta and the detuning $\Delta=\omega_{2}-\omega_{1}-\omega$, with $\hbar\omega_{1}$, $\hbar\omega_{2}$ are the energies of the uncoupled states $|1\rangle$ and $|2\rangle$, respectively, and $\omega$ is the frequency of the field mode. The system can be in two possible states $|i\rangle$, $i=1,2$ with $|1\rangle$ being the ground state of the system and $|2\rangle$ being the excited state, respectively. The four Paul operators are
\begin{align}
\begin{split}
1  & =|2 \rangle \langle 2|+ |1 \rangle  \langle 1|\\
 \hat{\sigma}_{3} & =|2 \rangle \langle 2|- | 1\rangle  \langle 1|\\
  \hat{\sigma}_{+}&= |2\rangle\langle 1 |\\
 \hat{\sigma}_{-}&= |1\rangle\langle 2 |.
\end{split}
\end{align}

Consider the initial state with the system in the ground state, viz.,
\begin{equation}
|\psi_{I}(0)\rangle=\sum_{n=0}^{\infty} a_{n}|1\rangle |n\rangle
\end{equation}
The initial state evolves with time according to
\begin{equation}
|\psi_{I}(t)\rangle=\sum_{n=0}^{\infty} \big{[}c_{1,n}(t)|1\rangle |n\rangle+ c_{2,n}(t)|2\rangle |n\rangle\big{]},
\end{equation}
where \cite{BR97}
\begin{align}
\begin{split}
c_{1,n}(t) & = a_{n}\big{(}\cos[\Omega_{R}(n) t/2] +i\frac{\Delta}{\Omega_{R}(n)}\sin[\Omega_{R}(n)t/2]\big{)}\\
c_{2,n}(t) & =-a_{n}\frac{2\lambda\sqrt{n}}{\Omega_{R}(n)}\sin[\Omega_{R}(n)t/2],
\end{split}
\end{align}
and $\Omega_{R}(n)= \sqrt{\Delta^2 +4 \lambda^2n}$.
We are interested in the probability of the system being in the ground state $|1\rangle$ at time $t=0$ for the resonant case ($\Delta=0$) and so
\begin{equation}
P(t)= \sum_{0}^{\infty}|c_{1,n}(t)|^2=\frac{1}{2}\sum_{n=0}^{\infty} |a_{n}|^2[1+\cos(2\lambda\sqrt{n}t)].
\end{equation}

\section{collapse and revival}
We consider the collapse and revival of initial states with one or more coherent or squeezed coherent photons. The dynamics is governed by the Hamiltonian (25). Consider first the initial states with coherent photons.

\subsection{n-photon coherent initial states}
The following $n$-photon coherent states, $n=1,2,3$, follow from (10),

\begin{equation}
|0,\alpha,1\rangle= e^{-|\alpha|^2/2}\sum_{n=0}^{\infty} \frac{|\alpha|^{n-1}}{\sqrt{n!}}(n-|\alpha|^2)|n \rangle,
\end{equation}
\begin{equation}
|0,\alpha,2\rangle= e^{-|\alpha|^2/2}\sum_{n=0}^{\infty} \frac{|\alpha|^{n-2}}{\sqrt{2! n!}}\Big{(}n(n-1)-2n|\alpha|^2+|\alpha|^4 \Big{)}|n \rangle,
\end{equation}
and
\begin{equation}
|0,\alpha,3\rangle= e^{-|\alpha|^2/2}\sum_{n=0}^{\infty} \frac{|\alpha|^{n-3}}{\sqrt{3!n!}}\Big{(}n(n-1)(n-2)-3n(n-1)|\alpha|^2+3n|\alpha|^4  -|\alpha|^6 \Big{)}|n \rangle.
\end{equation}

\begin{figure}
\begin{center}
\includegraphics[scale=0.3]{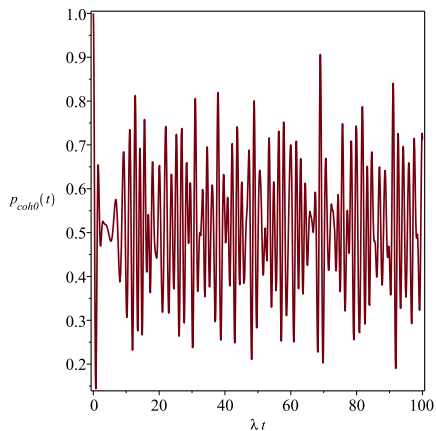}
\end{center}
\label{fig:theFig}
\caption{Plot for the probability $p_{coh0}(t)$ for the collapse and revival given in Eq. (30) with $|\alpha|=2$ in (11), that is, the zero-photon coherent state with four photons.}
\end{figure}
\begin{figure}
\begin{center}
\includegraphics[scale=0.3]{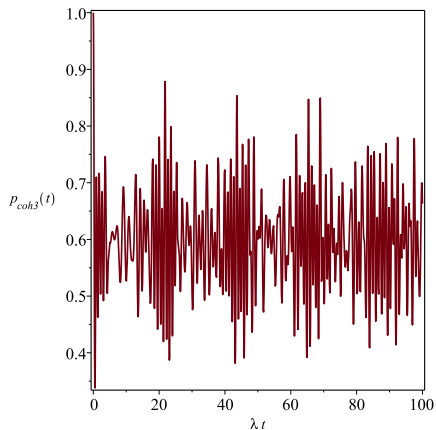}
\end{center}
\label{fig:theFig}
\caption{Plot for the probability $p_{coh3}(t)$ for the collapse and revival of the three-photon coherent state given by Eq. (33) with $|\alpha|=2$ and seven photons. }
\end{figure}

 We compare the collapse and revival of the three-photon coherent state (33) with that of the zero-photon coherent state (11).  We plot in Fig. 1 the collapse and revival of the zero-photon coherent state for
 $|\alpha|=2$, that is, the state with four photons. It is clear that we are considering states with low number of photons in order to compare their behavior with the three-photon coherent state (33), that is, seven photons.
 Fig. 2 shows the collapse and revival for the three-photon coherent state. Note the sharpening of the collapse and revival in Fig. 2 as contrasted to that of Fig. 1.

\begin{figure}
\begin{center}
\includegraphics[scale=0.3]{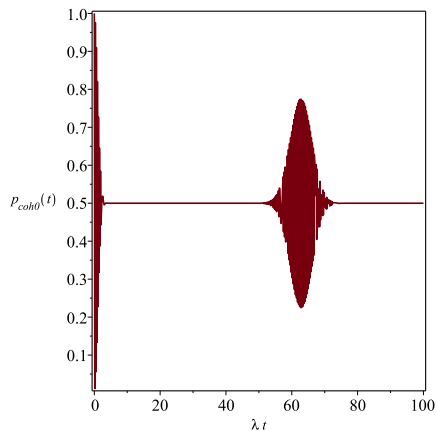}
 \end{center}
\label{fig:theFig}
\caption{Plot for the probability $p_{coh0}(t)$ for the collapse and revival given in Eq. (11) with $|\alpha|=10$, that is, the zero-photon coherent state with one hundred photons. }
\end{figure}

\begin{figure}
\begin{center}
\includegraphics[scale=0.3]{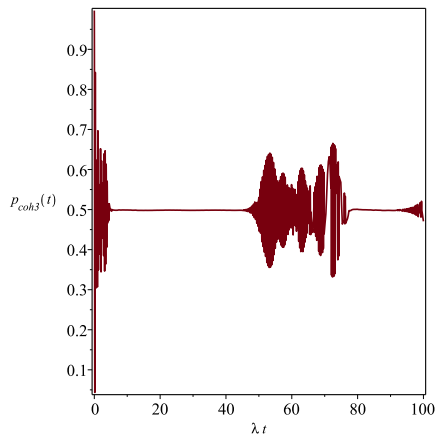}
\end{center}
\label{fig:theFig}
\caption{Plot for the probability $p_{coh3}(t)$ for the collapse and revival for the three-photon coherent state given in Eq. (33) with $|\alpha|=10$, that is, one hundred and three photons. }
\end{figure}

 On the other hand, one would expect that for a large number of photons, the collapse and revival behavior of the zero-photon coherent and the three-photon coherent states would be rather similar. However, that seems not to be the case. Fig. 3 show the probability of collapse and revival for the zero-photo coherent state given by (11) for $|\alpha|=10$, that is, one hundred photons.  The plot in Fig. 4 represents the collapse and
 survival of the three-photon coherent state with  $|\alpha|=10$ and one hundred and three photons. Therefore, the effect of a mere three photons in the three-photon coherent state is quite significant even for a difference of only $3\%$ in the numbers of photons between the two states.

\subsection{n-photon squeezed coherent initial states}

Here we consider the two-photon squeezed coherent states and minimize the quotient of the photon number variance (24) to the mean number of photons (19), viz., $\Delta n^2/\langle\hat{n}\rangle$. This is done in order to obtain optimal squeezing of the coherent state \cite{MA19}. One obtains for the value of $|\alpha|$

\begin{equation}
   |\alpha|^2 =-\frac{57}{80}+\frac{21}{80}e^{8r}-\frac{41}{20}e^{4r}+\frac{1}{2}e^{2r}  +\frac{1}{80}\sqrt{441e^{16r}-5488e^{12r}+560e^{10r} +31502e^{8r}-13120e^{6r}+13296e^{4r}-3440e^{2r}+1849}.
\end{equation}

It should be remarked that one has to be sure that the number of photons obtained from Eq. (19) turns out to be an integer. Accordingly, one has to choose the proper $r$ so that together with the $|\alpha|$
given by Eq. (34) gives rise to a whole number when substituted in Eq. (19).

As done previously for the $n$-photon coherent states in analyzing the collapse and revival, we now do the same for the $n$-photon squeezed coherent states.  Figs. 5 and 6 consider the case of low photon numbers whereas
Figs. 7 and 8 deal with high photon numbers. We observe precisely the same behavior as for the $n$-photon coherent states. Again, it is interesting that the effect in this case of two extra photons in the two-photon
squeezed coherent state has a significant effect in the collapse and revival for both low and high number of photons.

\begin{equation}
|\zeta,\alpha,0 \rangle= e^{-a^2(1-\tanh(r))/2}\sum_{n=0}^{\infty}\frac{   (\tanh(r))^{n/2} }{\sqrt{2^{n} n! \cosh(r)}} H_{n}(\frac{a}{\sqrt{2\sinh(r)\cosh(r)}}) |n\rangle
\end{equation}
\begin{align}
\begin{split}
|\zeta,\alpha,1 \rangle&= e^{-a^2(1-\tanh(r))/2}\sum_{n=0}^{\infty}\frac{   (\tanh(r))^{n/2} }{\sqrt{2^{n} n! \cosh(r)}} \Big{(}-\frac{1}{\sqrt{2 \sinh(r)\cosh(r)}} H_{n+1}(\frac{a}{\sqrt{2\sinh(r)\cosh(r)}})\\
& +a (1/\tanh(r) -1) H_{n}(\frac{a}{\sqrt{2\sinh(r)\cosh(r)}})                        \Big{)} |n\rangle
\end{split}
\end{align}
where $H_{n}(x)$ are the Hermite polynomials \cite{GR80} and  we have set $\varphi=2\theta$ and discarded an overall phase factor and
\begin{align}
\begin{split}
|\zeta,\alpha,2 \rangle & = e^{-a^2(1-\tanh(r))/2} \sum_{n=0}^{\infty} \frac{\tanh(r)^{n/2} }{\sqrt{2}\sqrt{2^{n} n! \cosh(r)}} \Big{(}\frac{a}{\sqrt{2\sinh(r)\cosh(r)}}\big{(}2-\tanh(r)-1/\tanh(r)\big{)}  H_{n+1}(\frac{a}{\sqrt{2\sinh(r)\cosh(r)}})\\
&   +   \big{(}n(\tanh(r)-1/\tanh(r))+\tanh(r)+a^2 (1/\tanh(r)-1)^2  \big{)} H_{n}(\frac{a}{\sqrt{2\sinh(r)\cosh(r)}} ) \Big{)}|n\rangle.
\end{split}
\end{align}

\begin{figure}
\begin{center}
\includegraphics[scale=0.3]{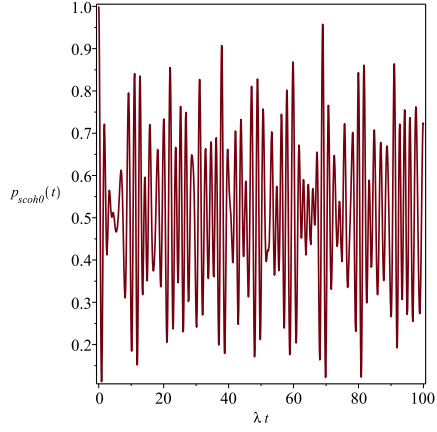}
\end{center}
\label{fig:theFig5}
\caption{Plot for the probability $p_{scoh0}(t)$ for the collapse and revival of the zero-photon squeezed coherent state with $|\alpha|=2.56230$ and $r=0.424875$ with three photons with the aid of (19). }
\end{figure}
\begin{figure}                                                                                                                                                                                                                                         \begin{center}
\includegraphics[scale=0.3]{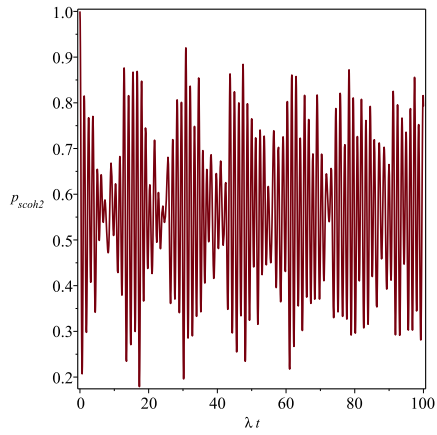}
\end{center}
\label{fig:theFig6}
\caption{Plot for the probability $p_{scoh2}(t)$ for the collapse and revival of the two-photon squeezed coherent state given in Eq. (37) with $|\alpha|=2.18536$ and $r=0.424875$ and so by (19), five photons. }
\end{figure}

\begin{figure}
\begin{center}
\includegraphics[scale=0.3]{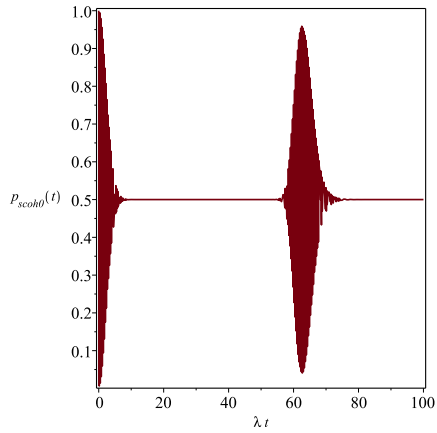}
\end{center}
\label{fig:theFig5}
\caption{Plot for the probability $p_{scoh0}(t)$ for the collapse and revival of the zero-photon squeezed coherent state given by Eq. (35) with  with $|\alpha|=24.4485$ and $r=0.8992$ and so by (19), one hundred photons. }
\end{figure}

\begin{figure}
\begin{center}
\includegraphics[scale=0.3]{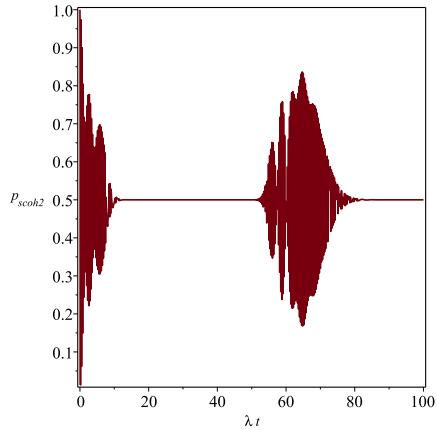}
\end{center}
\label{fig:theFig6}
\caption{Plot for the probability $p_{scoh2}(t)$ for the collapse and revival of the two-photon squeezed coherent state given by Eq. (37) with $|\alpha|=23.92344$ and $r=0.8992$ and so by (19), one hundred and two photons. }
\end{figure}

\newpage
\section{summary and conclusion}

We have introduced what we believe may be a new basis of the Hilbert space of photons, viz., the $n$-photon squeezed coherent states. We consider such states as initial conditions for the dynamics generated by the Jaynes-Cummings model in order to study the collapse and revival of these new states. It is curious that the difference in behavior between the zero-photon coherent state and the three-photon coherent state are quite significant for both low and high number of photons. The surprising case, of course, is that of the high number of photons where one would expect similar collapse and revival. This is also the case for the zero-squeezed
coherent states and the two-photon squeezed coherent states.

\newpage

\bibliography{basename of .bib file}

\end{document}